\documentclass[conference]{IEEEtran}
\IEEEoverridecommandlockouts
\usepackage{cite}
\usepackage{float} 
\usepackage{amsmath, amssymb, amsfonts}
\usepackage{algorithmic}
\usepackage{graphicx}
\usepackage{textcomp}
\usepackage{xcolor}
\usepackage{booktabs} 
\usepackage{graphicx}
\usepackage{epstopdf}
\def\BibTeX{{\rm B\kern-.05em{\sc i\kern-.025em b}\kern-.08em
    T\kern-.1667em\lower.7ex\hbox{E}\kern-.125emX}}
\begin{document}
\title{LLM-BSCVM: An LLM-Based Blockchain Smart Contract Vulnerability Management Framework}

\author{
\IEEEauthorblockN{
Yanli Jin\textsuperscript{1},
Chunpei Li\textsuperscript{1},
Peng Fan\textsuperscript{1},
Peng Liu\textsuperscript{1},
Xianxian Li\textsuperscript{1},
Chen Liu\textsuperscript{2},
Wangjie Qiu\textsuperscript{3}
}
\IEEEauthorblockA{
\textsuperscript{1}Guangxi Normal University, China \\
Email: jyl@stu.gxnu.edu.cn, licp@gxnu.edu.cn \\
\textsuperscript{2}Zhongguancun Laboratory, China \\
\textsuperscript{3}Beihang University, China
}
}
\maketitle
\begin{abstract}
Smart contracts are a key component of the Web 3.0 ecosystem, widely applied in blockchain services and decentralized applications. However, the automated execution feature of smart contracts makes them vulnerable to potential attacks due to inherent flaws, which can lead to severe security risks and financial losses, even threatening the integrity of the entire decentralized finance system. Currently, research on smart contract vulnerabilities has evolved from traditional program analysis methods to deep learning techniques, with the gradual introduction of Large Language Models. However, existing studies mainly focus on vulnerability detection, lacking systematic cause analysis and Vulnerability Repair. To address this gap, we propose LLM-BSCVM, a Large Language Model-based smart contract vulnerability management framework, designed to provide end-to-end vulnerability detection, analysis, repair, and evaluation capabilities for Web 3.0 ecosystem. LLM-BSCVM combines retrieval-augmented generation technology and multi-agent collaboration, introducing a three-stage method of “Decompose-Retrieve-Generate." This approach enables smart contract vulnerability management through the collaborative efforts of six intelligent agents, specifically: vulnerability detection, cause analysis, repair suggestion generation, risk assessment, vulnerability repair, and patch evaluation. Experimental results demonstrate that LLM-BSCVM achieves a vulnerability detection accuracy and F1 score exceeding 91\% on benchmark datasets, comparable to the performance of state-of-the-art (SOTA) methods, while reducing the false positive rate from 7.2\% in SOTA methods to 5.1\%, thus enhancing the reliability of vulnerability management. Furthermore, LLM-BSCVM supports continuous security monitoring and governance of smart contracts through a knowledge base hot-swapping dynamic update mechanism. It not only provides developers with comprehensive vulnerability management services but also effectively improves the overall security of the Web 3.0 ecosystem. To support the development of Web 3.0 and blockchain security research communities, the code for this framework is open-source and available at https://github.com/sosol717/LLM-BSCVM.
\end{abstract}

\begin{IEEEkeywords}
Web 3.0, Blockchain, Smart Contract, Vulnerability Management, Large Language Model (LLM)
\end{IEEEkeywords}

\section{Introduction}
With the rapid development of Web 3.0 and blockchain technology, smart contracts have become a crucial foundation for decentralized applications and the decentralized finance ecosystem. However, due to the automated execution and immutability of smart contracts, once vulnerabilities exist, attackers can exploit the logical flaws in the code, leading to severe security threats. For example, the 2016 DAO attack resulted in a loss of \$60 million \cite{ghaleb2018addressing}, and in 2018, the BEC contract caused the value of tokens to drop to zero due to an integer overflow vulnerability \cite{hessenauer2018batch}. According to statistics, smart contract vulnerabilities have led to an accumulated economic loss of over \$20 billion \cite{sayeed2020smart}. Therefore, comprehensive vulnerability detection, cause analysis, repair suggestion generation, and risk assessment before deploying smart contracts are crucial for ensuring the security of the Web 3.0 ecosystem.

Several methods have been proposed for smart contract vulnerability detection, primarily including the following:\textbf{ (1) Traditional Methods:} such as formal verification \cite{kalra2018zeus}, symbolic execution \cite{luu2016making}, and fuzz testing \cite{jiang2018contractfuzzer}, These methods rely on expert knowledge and can provide some vulnerability detection capabilities. However, they tend to have a high false positive rate when applied to complex contracts and are difficult to scale \cite{liu2023vulnerable}. \textbf{(2) Deep Learning Methods (e.g., RNN, LSTM, CNN \cite{chen2023diversevul}, \cite{wang2020contractward}, \cite{zhen2024gnn}):} These methods can automatically learn vulnerability features in contracts, reducing human intervention. However, they are typically limited to pattern matching for vulnerabilities and lack a deep understanding of code syntax and semantics, making it challenging to generate detailed vulnerability cause analysis and repair solutions.\textbf{ (3) Generative Large Language Model (LLM) Methods:} Recent research has shown that LLMs possess a deep understanding of code and can capture long-distance dependencies within the code, demonstrating strong generalization capabilities in vulnerability detection \cite{wang2023review}. However, current research on LLMs for smart contract vulnerabilities is still limited to the detection stage. There are still significant shortcomings in the areas of vulnerability explainability analysis and automated repair \cite{sun2024gptscan}, \cite{ma2024combining}, \cite{hu2023large}.

The research in \cite{chu2023survey}, \cite{kumar2024vulnerabilities}indicates that smart contract vulnerabilities are a multidimensional issue that requires comprehensive exploration. However, existing smart contract vulnerability detection methods mainly focus on vulnerability identification, while lacking fine-grained cause analysis and automated repair suggestions \cite{sun2024gptscan}, \cite{hu2023large}, \cite{liu2018reguard}. This results in detection outcomes that are not directly actionable for contract repair. Furthermore, most methods adopt black-box outputs, making it difficult for developers to understand the root causes of vulnerabilities. The repair process still relies on expert knowledge and manual analysis, which decreases the efficiency and scalability of vulnerability fixes \cite{chen2023diversevul}, \cite{wang2020contractward}, \cite{zhen2024gnn}. Although large models have demonstrated strong code comprehension capabilities in the field of vulnerability detection \cite{ma2024combining}, \cite{hu2023large}, current research lacks a systematic vulnerability management framework, making it difficult to cover the post-detection processes of analysis, evaluation, and repair. Therefore, building an end-to-end smart contract vulnerability management system that encompasses vulnerability detection, cause analysis, repair suggestion generation, risk assessment, and audit reporting, in order to enhance the automation, explainability, and reliability of vulnerability governance, remains a core challenge that needs to be addressed in the current research landscape. Research \cite{liu2024exploring} further shows that large language models possess great potential in handling vulnerability management tasks, providing new insights for developing interpretable vulnerability management methods and laying the technological foundation for more comprehensive vulnerability management services.

To address the aforementioned issues, this paper proposes LLM-BSCVM, a Large Language Model (LLM)-based smart contract vulnerability management framework, which provides comprehensive capabilities for vulnerability detection, analysis, repair, and evaluation. To build a more robust vulnerability management system, LLM-BSCVM introduces a “Decompose-Retrieve-Generate" three-stage approach specifically for smart contract vulnerability management. This approach aims to systematically address the detection, analysis, and repair of smart contract vulnerabilities, enhancing the automation, explainability, and reliability of vulnerability governance.
LLM-BSCVM employs a three-stage vulnerability management method:\\
\indent(1) Task Decomposition: Based on the concept of multi-agent collaboration \cite{chen2023agentverse}, the vulnerability management task is subdivided into six sub-tasks: vulnerability detection, cause analysis, repair suggestion generation, risk assessment, vulnerability repair, and patch evaluation. Each sub-task is independently handled by different agents, and multiple agents collaborate, with the results of preceding tasks supporting subsequent ones, forming a progressive inference process.\\
\indent(2) Knowledge Retrieval: By integrating retrieval-augmented generation \cite{lewis2020retrieval}, during the execution of each agent's task, the vulnerability knowledge base and external data sources are dynamically retrieved to enhance the model's understanding of vulnerability causes and repair strategies, improving the accuracy of inference.\\
\indent(3) Result Generation: The agents, in conjunction with the retrieved relevant knowledge, generate vulnerability detection reports, cause analyses, repair suggestions, and final audit reports, thus improving the explainability and automation of vulnerability management.

In the vulnerability detection phase, we use a separately fine-tuned detection model (CodeLlama) to provide initial results, ensuring detection accuracy. Subsequent tasks are progressively inferred by the foundational large model (CodeLlama), and finally, after vulnerability repair is completed, a more computationally expensive advanced LLM is used for final audit evaluation to ensure the reliability of the output results. The main contributions of this paper are as follows:
\begin{itemize}
\item LLM-BSCVM: The first smart contract vulnerability management framework, integrating vulnerability detection, cause analysis, risk assessment, vulnerability repair, repair verification, and report generation, thereby constructing a complete vulnerability governance system suitable for the Web 3.0 ecosystem.
\item Decompose-Retrieve-Generate Method: The proposed “Decompose-Retrieve-Generate" three-stage method for smart contract vulnerability management, combining multi-agent collaboration and retrieval-augmented generation. This method enhances LLM’s explainability and accuracy in smart contract vulnerability management through task decomposition, dynamic knowledge expansion, and inference enhancement.
\item Experimental Validation: Experiments on a smart contract dataset validate the effectiveness of LLM-BSCVM. The results show that the framework's vulnerability detection accuracy and F1 score exceed 91\%, comparable to state-of-the-art (SOTA) methods. At the same time, the false positive rate is reduced from 7.2\% in SOTA methods to 5.1\%, significantly decreasing the error alarm rate and improving the precision and feasibility of vulnerability repair.
\end{itemize}
\hspace{\parindent}This study not only advances the application of AI in smart contract vulnerability management but also offers a novel approach to Web 3.0 and blockchain security governance. The code is open-sourced (https://github.com/sosol717/LLM-BSCVM) to support the development of the Web 3.0 and blockchain security research community.

\section{Related work}
\subsection{Smart Contract Vulnerability Detection}
Traditional methods are widely used for detecting vulnerabilities in smart contracts, including fuzz testing, symbolic execution tools, and formal verification methods. Fuzz testing tools such as Contractfuzzer \cite{jiang2018contractfuzzer}, Reguard \cite{liu2018reguard}, and Soliaudit \cite{liao2019soliaudit} discover vulnerabilities by simulating various inputs at runtime. However, these tools typically only cover a subset of code paths, potentially missing latent vulnerabilities, and often have a high false positive rate. Symbolic execution tools like Oyente \cite{luu2016making}, Manticore \cite{mossberg2019manticore}, and WANA \cite{jiang2021wana} conduct boundary checks by analyzing the execution paths of contracts. While they can identify more complex vulnerabilities, they are computationally expensive and are often limited to smaller or simpler contracts. Formal verification tools, such as Zeus \cite{kalra2018zeus} and VeriSmart \cite{so2020verismart}, provide rigorous mathematical proofs, but their comprehensive verification of complex contracts remains challenging.

With the development of deep learning technologies, researchers have started using deep learning models to learn vulnerability features from contract samples for detection. Diversevul \cite{chen2023diversevul} employs deep learning algorithms to extract vulnerability features from contract samples, and then uses neural network models to detect different types of vulnerabilities. Contractward \cite{wang2020contractward} analyzes the bytecode and opcodes of contracts, leveraging deep neural network models to detect potential vulnerabilities, including common issues like reentrancy attacks. DA-GNN \cite{zhen2024gnn}, on the other hand, constructs a contract's control flow graph (CFG) and combines graph-based features to capture the complex relationships between nodes within the contract, thereby more accurately identifying and detecting vulnerabilities in smart contracts.

In recent years, Large Language Models (LLMs) have provided new approaches for vulnerability detection. Researchers have not only utilized traditional deep learning models but also applied LLMs for vulnerability detection. GPTScan \cite{sun2024gptscan} attempts to combine large models with static analysis for detecting logical vulnerabilities in smart contracts. GPTLENS \cite{hu2023large} is a two-stage adversarial framework that uses GPT-4 to mine potential vulnerabilities in smart contracts, with the goal of identifying as many real vulnerabilities as possible. TrustLLM \cite{ma2024combining} conducts intuitive smart contract audits and generates audit explanations through majority voting and multiple-prompt fine-tuning. LLMSmartSec \cite{mothukuri2024llmsmartsec} leverages GPT-4 to understand smart contracts and trains an LLMGraphAgent to achieve low-cost automated security auditing. LLM4Vuln \cite{sun2024llm4vuln} accurately evaluates the performance of LLMs in vulnerability detection by separating the active search for additional information, employing relevant vulnerability knowledge, and generating structured results.

However, all of these studies primarily explore vulnerabilities from a single dimension and mainly focus on vulnerability detection tools, which presents two major limitations. On one hand, detection tools lack fine-grained interpretability analysis \cite{chen2023diversevul}, \cite{wang2020contractward}, \cite{zhen2024gnn}, failing to effectively reveal the root causes of vulnerabilities, and they do not provide reliable repair suggestions, leading developers to still rely heavily on manual analysis and expert experience during the vulnerability repair process. On the other hand, these tools do not cover the full lifecycle of vulnerabilities \cite{sun2024gptscan}, \cite{hu2023large}, especially in terms of interpretability analysis and automated repair, where there are significant shortcomings. This results in an inability to address diverse security needs, leading to issues such as inefficiency, insufficient accuracy, and increased security risks.

\subsection{Large Language Models (LLMs)}
Large Language Models (LLMs) are a class of deep learning models based on the Transformer architecture \cite{vaswani2017attention}. They undergo pre-training on large-scale text data through self-supervised learning, thereby acquiring rich language knowledge. As the training data increases, LLMs are capable of processing longer context information, demonstrating enhanced abilities in language understanding and generation. In addition to capturing subtle nuances in text, LLMs can also generate grammatically correct and semantically coherent natural language, improving the quality and fluency of text generation. In the domain of code understanding and generation, LLMs have undergone several evolutions. From the GPT series, which supports multilingual code understanding \cite{radford2019language}, to the specially optimized CodeLlama \cite{roziere2023code}, code-specific large language models have shown tremendous potential in program understanding, analysis, and generation.

Fine-tuning is the process of domain-specific optimization of a pre-trained large language model. During the pre-training phase, LLMs learn extensive language knowledge by training on large-scale general datasets. However, for certain specific tasks or domains, the pre-trained model may not be fully applicable. Through fine-tuning, LLMs can be optimized for specific application scenarios, thereby improving their performance on targeted tasks. Compared to training from scratch, fine-tuning can significantly enhance model performance in a shorter time frame while effectively saving computational resources.

\subsection{Retrieval-Augmented Generation (RAG)}
Retrieval-Augmented Generation (RAG) \cite{lewis2020retrieval} introduces an external knowledge retrieval mechanism into the generation process of large language models, effectively addressing the knowledge gap that large models may have in specific domains. In traditional large pretrained language models, generation methods typically rely on the parameters learned during pretraining to generate text. However, this approach may produce inaccurate outputs when handling complex tasks or scenarios requiring real-time updates. The RAG technique enables the generation model to dynamically retrieve relevant documents from an external knowledge base while processing the input, merging the retrieved information with the input, thereby enhancing the accuracy and reliability of the generated results.\\

\section{Detailed Design of LLM-BSCVM}
\subsection{Framework Overview}
\begin{figure*}[h]
  \centering
  \includegraphics[width=\textwidth]{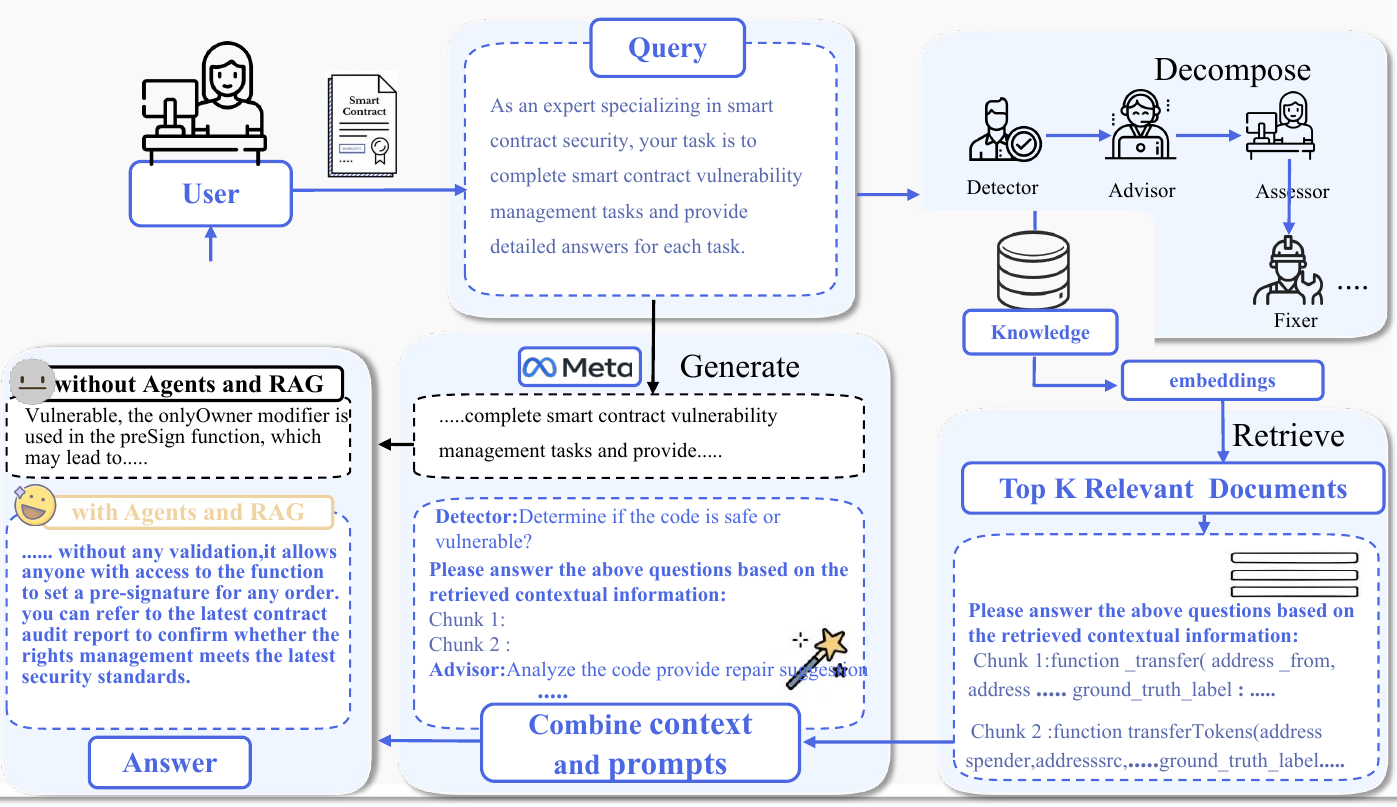}  
  \caption{Framework of our proposed approach LLM-BSCVM.}
  \label{fig:1}
\end{figure*}

The overall framework we propose is shown in Figure \ref{fig:1}. The LLM-BSCVM framework implements smart contract vulnerability management through a three-stage "Decompose-Retrieve-Generate" approach. This method organically combines retrieval-augmented generation (RAG) with a multi-agent collaborative task decomposition mechanism, breaking down the complex vulnerability management process into several subtasks. At each subtask, relevant specialized knowledge is dynamically retrieved, ensuring that the model's output of repair suggestions and evaluation results is based on reliable and accurate knowledge.
\hspace{\parindent}We will discuss each stage in more detail in the following chapters.
\begin{itemize}
\item \textbf{Task Decomposition Stage: }The vulnerability management task is subdivided into six subtasks, each handled independently by a different agent. Multiple agents work collaboratively, with the results of previous tasks providing support for subsequent tasks, forming a progressive reasoning process.
\item \textbf{Knowledge Retrieval Stage: }Each agent uses retrieval-augmented generation (RAG) technology to access relevant information in real-time from the vulnerability knowledge base and external data sources.
\item \textbf{Result Generation Stage: }The LLM integrates the knowledge obtained from the retrieval stage into the prompts and generates interpretable results for vulnerability detection, evaluation, analysis, and repair. Finally, an objective audit report is generated.
\end{itemize}

\subsection{Task Decomposition Stage}
Smart contract vulnerability management is a multi-step, complex process that covers various stages, including vulnerability detection, repair suggestions, risk assessment, and vulnerability repair. To ensure the efficient execution and accuracy of each task, we need to break the entire process down into multiple independent yet interrelated subtasks. Each subtask has its own clear objectives and responsibilities. While minimizing interference between tasks, these subtasks must also provide necessary support for subsequent tasks.\\
\begin{figure*}[h]
  \centering
  \includegraphics[width=\textwidth]{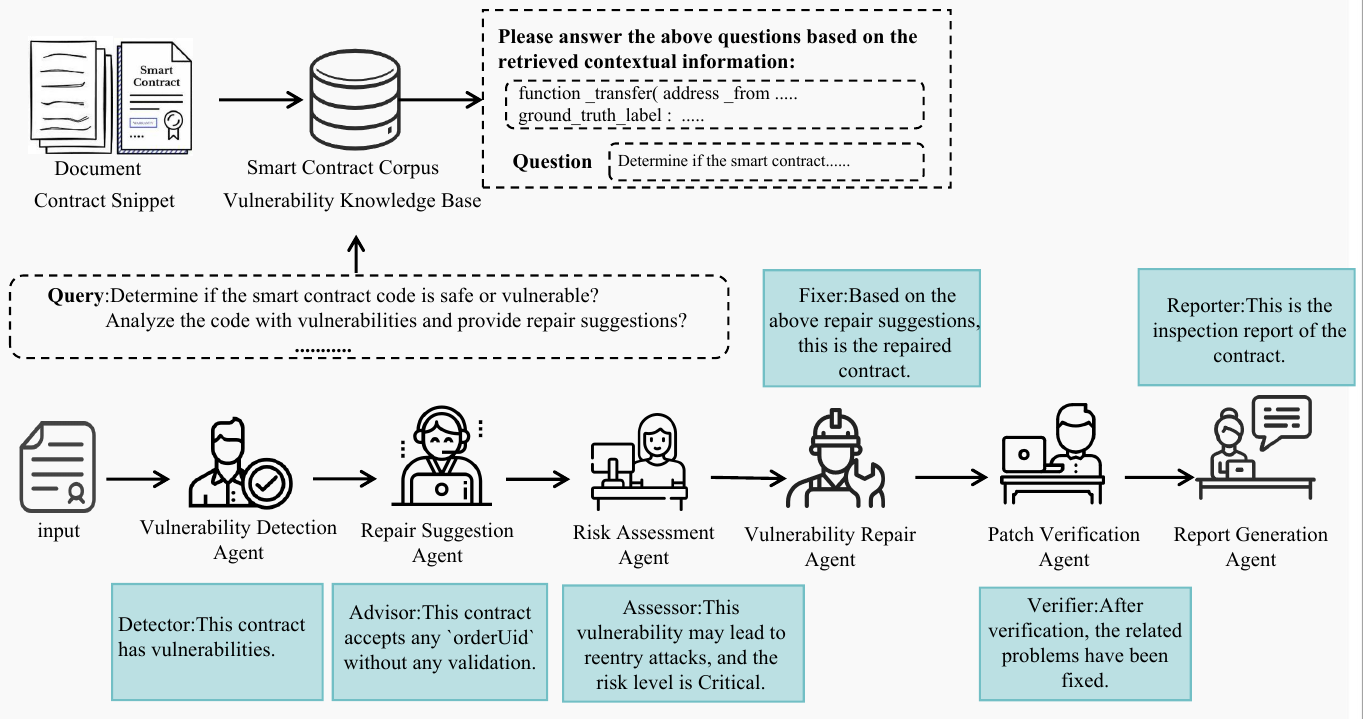}  
  \caption{The specific of task decomposition.}
  \label{fig:2}
\end{figure*}
\indent Multi-agent collaboration is an effective method for solving complex problems, particularly those that require multiple independent entities to work together. In the smart contract vulnerability management process, the entire workflow is broken down into six subtasks, each assigned to an independent agent. Each agent, based on its specific task objectives and expertise, performs independent reasoning and decision-making, passing its output to subsequent agents. This collaborative working mechanism ensures efficient task interconnection and information flow, thereby improving the overall accuracy and reliability of the management process. As shown in Figure \ref{fig:2}, the entire vulnerability management process is divided into six subtasks: vulnerability detection, repair suggestion generation, risk assessment, vulnerability repair, patch correctness evaluation, and detection report generation. Each subtask is handled by a different agent, corresponding to its specific task:\\
\indent\textbf{Vulnerability Detection Agent (Detector): }As the core of smart contract auditing, the vulnerability detection agent is responsible for identifying potential vulnerabilities in the contract and providing accurate detection results for subsequent tasks. To improve the accuracy of vulnerability detection, the results from three dimensions are integrated at this stage. (1) Static Analysis, Based on a predefined pattern library, common vulnerabilities, such as reentrancy attacks and arithmetic overflows, are detected; (2) Using Retrieval-Augmented Generation (RAG) technology, top-k contracts similar to the target contract are retrieved in real time, and relevant information is sourced from the contract library; (3) Inference Analysis:  A fine-tuned model deeply understands the business logic of the contract, assessing potential security issues.\\
\indent Each dimension’s analysis results are marked as “safe” or “vulnerable.” Based on this, we design two decision-making approaches to combine the results from all dimensions: one is through weighted fusion, combining a dynamic threshold mechanism to determine the final security; the other is through a voting mechanism, where the contract’s final security is decided by majority vote. It should be noted that the analysis results of the first two dimensions are not directly provided to the model for the final decision, as experiments show that excessive external information may introduce noise, affecting the accuracy of detection.\\
\indent\textbf{Repair Suggestion Agent (Advisor): }After vulnerability detection, the repair suggestion agent is responsible for providing targeted repair solutions for the detected vulnerabilities. Using RAG technology, it retrieves real-time information from the vulnerability knowledge base and, combined with the model’s generative capabilities, provides specific repair suggestions for each vulnerability. The repair solutions include root cause analysis of the vulnerability, potential impact assessment, repair steps, and preventive measures, ensuring the comprehensiveness and effectiveness of the repair solutions.\\
\indent\textbf{Risk Assessment Agent (Assessor): }The risk assessment agent systematically evaluates the risk level of each vulnerability by analyzing audit reports from major security audit agencies, vulnerability disclosure data, and the CVSS score standard from the CVE vulnerability database. Based on a four-level risk assessment system (Critical, High, Medium, Low), and leveraging the model’s reasoning capabilities, the agent assigns a risk level to each vulnerability, which provides a basis for prioritizing the subsequent repair tasks.\\
\indent\textbf{Vulnerability Repair Agent (Fixer): }The vulnerability repair agent is responsible for fixing the vulnerabilities in the smart contract based on the repair suggestions and risk assessment results. The agent first sorts vulnerabilities according to their repair priority, then, considering contextual information and dependencies, generates repair code that complies with programming standards, ensuring the security and effectiveness of the repair process.\\
\indent\textbf{Patch Verification Agent (Verifier): }The repaired code must undergo verification to ensure no new security issues are introduced. We adopt the concept of multi-agent debate, using independent evaluation models to verify the repaired code. The verification process mainly includes two aspects: first, ensuring that the repair successfully eliminates the vulnerability, and second, ensuring that no new security issues are introduced during the repair process.\\
\indent\textbf{Report Generation Agent (Reporter): }Finally, the report generation agent integrates the analysis results from the previous stages into a complete audit report. The report includes seven key sections: contract basic information overview, executive summary, audit methodology explanation, vulnerability discovery summary, in-depth analysis report, improvement suggestions, and compliance disclaimer, providing a comprehensive reference for smart contract developers.

\subsection{Knowledge Retrieval Stage}
\begin{figure}
  \centering
  \includegraphics[width=1\linewidth]{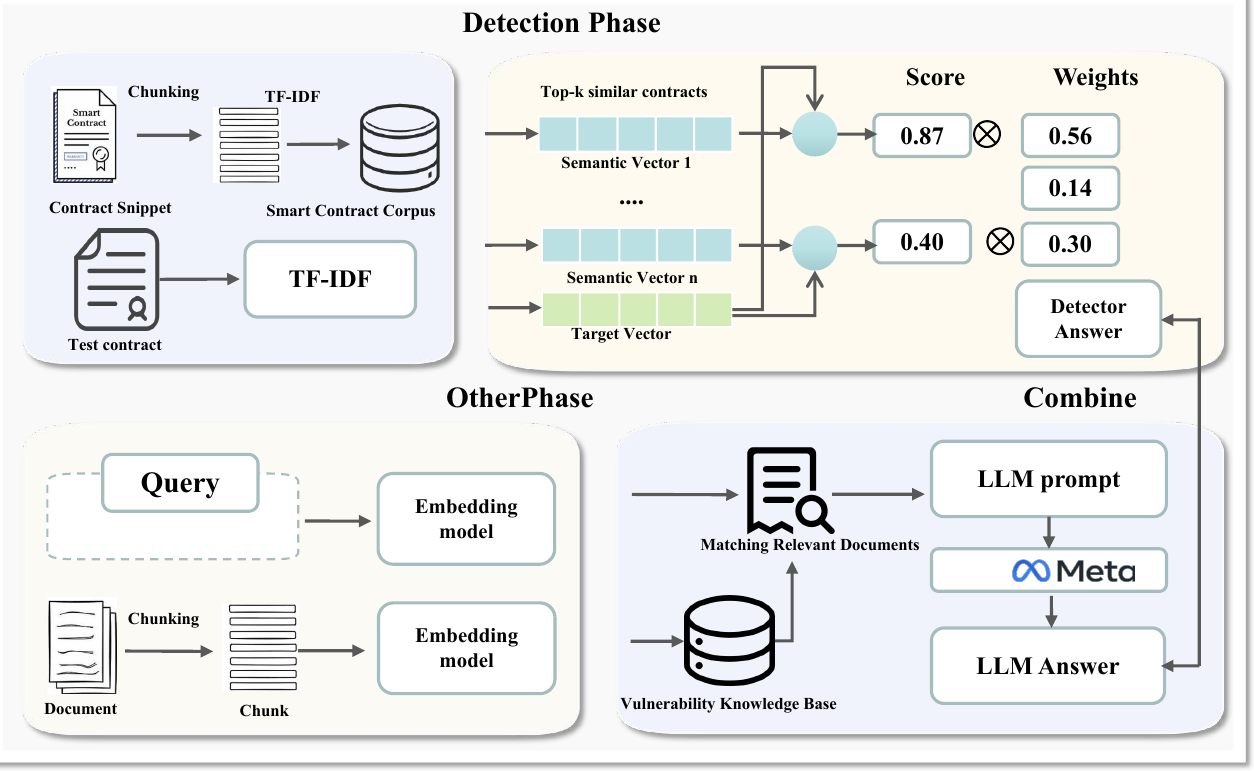}
  \caption{The specific of knowledge retrieval.}
  \label{fig:3}
\end{figure}

In terms of knowledge base construction, as shown in Figure \ref{fig:3}, we have utilized RAG technology to build two knowledge bases aimed at smart contract vulnerability management.
\begin{itemize}
\item Smart Contract Corpus: This corpus contains a large volume of smart contract code, primarily used for similarity retrieval in the vulnerability detection stage. The data is sourced from \cite{ma2024combining}, and collected from the well-known auditing website Solodit \cite{solodit2024}, analyzing a total of 263 smart contract audit reports.
\item Vulnerability Knowledge Base: This knowledge base stores documents related to smart contract vulnerabilities, including smart contract audit reports \cite{zheng2024dappscan} from renowned security institutions, security best practices \cite{riady2019bestpractices}, and programming standard documents \cite{eea2024ethtrust}. These documents are primarily sourced from leading companies in the industry, such as Solidit \cite{solodit2024} and Smart Contract Weakness Classification (SWC) \cite{swc-registry}. The vulnerability knowledge base provides relevant documents and background information for tasks at all stages.
\end{itemize}
\hspace{\parindent}For processing the smart contract corpus, we applied the TF-IDF algorithm to vectorize the contract code. By computing the product of Term Frequency (TF) and Inverse Document Frequency (IDF), we are able to capture the key features within the contract code. We then used cosine similarity to measure the similarity between the target contract and other contracts in the corpus, selecting the top-k most similar contracts for further analysis. To improve the accuracy of similarity retrieval, we assigned different weights to contracts based on their similarity rankings (the higher the rank, the greater the weight) and calculated the final probability of vulnerability presence through weighted computation. As shown in Formula 1, where Va and Vb represent the contract encoding vectors to be compared.\\
\indent For processing the vulnerability knowledge base data, we employed vector embedding techniques to convert unstructured text into vector representations in a high-dimensional semantic space. This enables LLM-BSCVM to perform semantic similarity retrieval based on the specific requirements of the analysis stage (e.g., vulnerability detection or repair recommendation generation), providing the model with specialized knowledge support.
\begin{equation}
\text{cosine similarity}(V_a, V_b) = \frac{V_a \cdot V_b}{\|V_a\| \|V_b\|}
\end{equation}
\subsection{Result Generation Stage}
In the process of smart contract vulnerability management, each intelligent agent retrieves and integrates relevant domain knowledge, embedding it into the task execution process. Based on this, it generates explainable vulnerability detection, evaluation, analysis, and repair results. Ultimately, based on the collaborative outcomes of all intelligent agents, a comprehensive audit report is produced, offering developers practical and feasible vulnerability repair strategies and improvement recommendations.

\begin{figure}[h]
  \centering
  \includegraphics[width=1\linewidth]{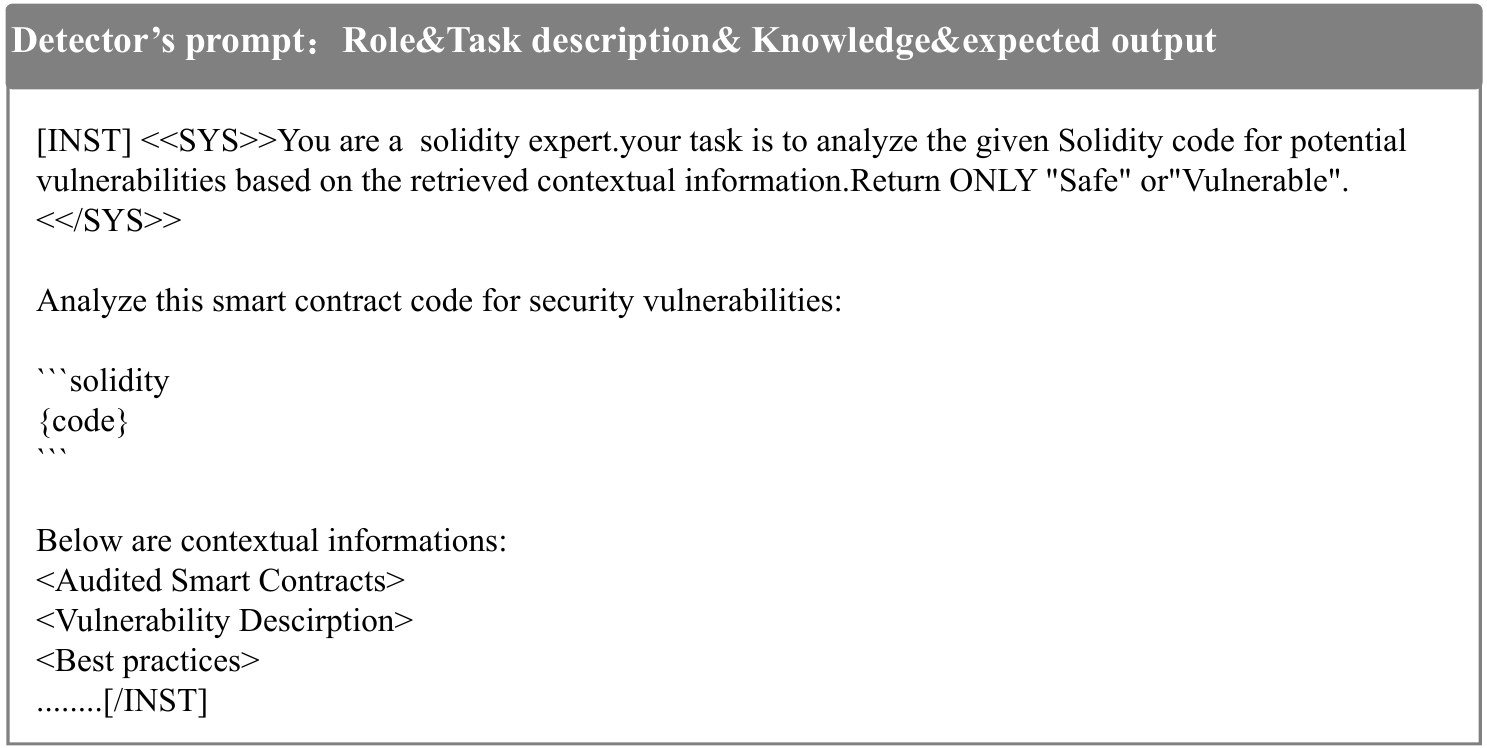}
  \caption{Prompt Template of LLM-BSCVM: An Example for Vulnerability Detection Tasks.}
  \label{fig:4}
\end{figure}

The basic prompt template consists of four core components: role-playing, task description, expected output, and background information. Taking vulnerability detection as an example, as shown in Figure \ref{fig:4}, the input in the template is the smart contract code to be analyzed, with “{code}" serving as a placeholder. The background information includes best security practices, Solidity programming guidelines, etc. This information is dynamically adjusted based on task requirements to help the model better understand the task objectives and provide structured feedback.
\section{Evaluation and Results}
\subsection{Experimental Setup}
Our dataset is compiled from two sources: \\
\indent(1) TrustLLM \cite{ma2024combining}, which originates from the renowned auditing website Solodit \cite{solodit2024}, comprising a total of 263 smart contract audit reports.\\
\indent(2) smart contract audit reports from Dappscan \cite{zheng2024dappscan}, which includes analysis of 1,199 open-source audit reports from 29 security teams.\\
\indent In this study, we set the value of k in similarity retrieval to 5, and we also designed a flexible interface to integrate various large language models (LLMs), including Codellama \cite{roziere2023code}, CodeBERT \cite{feng2020codebert}, CodeT5 \cite{wang2021codet5}, and Llama \cite{touvron2023llama}. Within our framework, we employed different model configurations according to task requirements: the vulnerability detection task uses Codellama fine-tuned with LoRA; tasks such as vulnerability fix suggestion generation, risk level assessment, vulnerability repair, and report generation use the base Codellama model. For patch correctness evaluation, we adopted the multi-agent debate concept \cite{wang2024contracttinker}, introducing GPT-4 as an independent verifier to assess whether the patch successfully fixed the target vulnerability.
\subsection{Experimental Results}
\textbf{Vulnerability Detection. }To evaluate the performance of LLM-BSCVM in vulnerability detection, we employed accuracy, precision, recall, and F1-score as evaluation metrics. The experiments were conducted using the same dataset as TrustLLM. LLM-BSCVM (W) refers to the weighted fusion approach, with weight distribution as follows: model-based detection (70\%), static analysis (10\%), and retrieval-based detection (20\%). In contrast, LLM-BSCVM (V) adopts a majority voting mechanism, making decisions based on model predictions, static analysis results, and similarity matching outcomes.\\
\indent As shown in Table ~\ref{tab:performance1}, the weighted approach, LLM-BSCVM (W), achieved the highest detection accuracy of 91.11\% and precision of 94.95\%. These results indicate that LLM-BSCVM (W) effectively identifies and localizes vulnerabilities in smart contracts, demonstrating strong detection capability. Comparatively, although LLM-BSCVM (V) also performed well with an F1-score of 89\%, recall of 86\%, precision of 93\%, and accuracy of 89\%, it exhibited slightly inferior performance. We hypothesize that this difference arises due to the majority voting method employed in LLM-BSCVM (V), which aggregates results from different components through a simple voting mechanism. However, this approach may amplify the influence of weaker components (such as static analysis or retrieval), negatively impacting the final decision and leading to a slight decrease in overall performance.In contrast, LLM-BSCVM (E) incorporates retrieved similar contracts and security documentation as contextual information to theoretically enhance detection accuracy. However, the actual results show a decline in performance. We speculate that the introduction of excessive contextual information may have distracted the model’s attention, thereby reducing its ability to accurately detect specific vulnerabilities.\\
\begin{table}[htbp]
    \centering
    \caption{Comparison of Detection Performance of LLM-BSCVM Using Different Methods}
    \label{tab:performance1}
    \begin{tabular}{lcccc}
        \toprule
        Approach & F1 & Recall & Precision & Accuracy \\
        \midrule
        LLM-BSCVM(E) & 0.7890 & 0.7125 & 0.8467 & 0.8042 \\
        LLM-BSCVM(V) & 0.8996 & 0.8689 & 0.9326 & 0.8999 \\
        LLM-BSCVM(W) & 0.9104 & 0.8743 & 0.9506 & 0.9111 \\
        \bottomrule
    \end{tabular}
\end{table}
\indent Additionally, as presented in Table ~\ref{tab:performance2}, LLM-BSCVM achieves performance comparable to TrustLLM \cite{ma2024combining} in terms of F1-score and precision, reaching high scores of 91\% and 91\%, respectively. An F1-score of 91\% indicates that the model maintains a strong balance between precision and recall in vulnerability detection. A precision of 91\% suggests that 91\% of the contracts predicted as vulnerable are indeed true positives. These metrics demonstrate that LLM-BSCVM effectively distinguishes between vulnerable and non-vulnerable contracts, ensuring both efficiency and accuracy in detection.\\
\indent Regarding the false positive rate, LLM-BSCVM achieved a false positive rate of 5.1\%, which is 2.2 percentage points lower than TrustLLM’s 7.2\%. The false positive rate (FPR) is a crucial indicator of a model’s detection accuracy, where a lower FPR implies a more precise identification of normal contracts, reducing the risk of misclassification. By integrating a fine-tuned large language model, static analysis, and retrieval-augmented techniques, LLM-BSCVM identifies and verifies vulnerabilities at multiple levels, mitigating errors caused by insufficient knowledge coverage and significantly reducing the false positive rate.

We further compared LLM-BSCVM with both base models and fine-tuned models. The base models include Codellama 13B, Codellama 7B, CodeBERT, CodeT5, and Llama, while the fine-tuned models consist of their respective fine-tuned versions. As shown in Table ~\ref{tab:performance2}, LLM-BSCVM outperforms all baseline models across all evaluation metrics, particularly in accuracy, where it surpasses Codellama 13B by approximately 48 percentage points. This result suggests that base models (e.g., Codellama and CodeBERT) lack sufficient contextual understanding and domain knowledge when handling complex smart contracts, leading to significantly lower accuracy compared to LLM-BSCVM.
\begin{table}[htbp]
    \centering
    \caption{Comparison of Detection Performance between LLM-BSCVM and Zero-Shot Learning LLMs}
    \label{tab:performance2}
    \begin{tabular}{lcccc}
        \toprule
        Approach & F1 & Recall & Precision & Accuracy \\
        \midrule
        Codellama 7B & 0.5278 & 0.6749 & 0.4333 & 0.3766 \\
        Codellama 13B & 0.5791 & 0.8708 & 0.4338 & 0.4255 \\
        CodeT5 & 0.6183 & 0.7568 & 0.5226 & 0.5176 \\
        CodeBERT & 0.5208 & 0.5464 & 0.4975 & 0.4810 \\
        Ilama 8B & 0.5938 & 0.8087 & 0.4691 & 0.4288 \\
        LLM-BSCVM(W) & 0.9104 & 0.8743 & 0.9506 & 0.9111 \\
        \bottomrule
    \end{tabular}
\end{table}

As illustrated in Table Table ~\ref{tab:performance_lora}, fine-tuned models exhibit substantial performance improvements over their base counterparts across all metrics, particularly in precision, recall, and F1-score, highlighting their enhanced capability in vulnerability detection. For instance, the fine-tuned versions of Codellama 13B and CodeBERT show significant improvements in precision and recall, indicating that the fine-tuning process enhances the model’s understanding of smart contract contexts and its ability to accurately localize vulnerabilities. However, despite these improvements, LLM-BSCVM consistently outperforms all fine-tuned models, particularly in accuracy, where even the fine-tuned versions of Codellama 13B and CodeBERT fall short of LLM-BSCVM’s performance.

\hspace*{\parindent}Furthermore, to validate the effectiveness of each component in the vulnerability detection method, we conducted an ablation study by systematically removing different components to evaluate their contributions to overall performance. The LLM-BSCVM vulnerability detection framework consists of three core components: (1) a LoRA fine-tuned CodeLlama-13B model, (2) a static analysis module based on predefined vulnerability patterns, and (3) a retrieval-augmented module leveraging historical vulnerability knowledge.
\begin{table}[htbp]
    \centering
    \caption{Comparison of Detection Performance between LLM-BSCVM and LoRA Fine-Tuned LLMs}
    \label{tab:performance_lora}
    \begin{tabular}{lcccc}
        \toprule
        Approach & F1 & Recall & Precision & Accuracy \\
        \midrule
        Codellama 7B(Lora) & 0.8451 & 0.8445 & 0.9211 & 0.8954 \\
        Codellama 13B(Lora) & 0.8918 & 0.8661 & 0.9388 & 0.9027 \\
        CodeT5(Lora) & 0.8543 & 0.7887 & 0.9411 & 0.8543 \\
        CodeBERT(Lora) & 0.8121 & 0.7230 & 0.9111 & 0.8564 \\
        Ilama 8B(Lora) & 0.8231 & 0.7554 & 0.9233 & 0.8422 \\
        TrustLLM & 0.9121 & 0.8934 & 0.9316 & 0.9111 \\
        LLM-BSCVM(W) & 0.9104 & 0.8743 & 0.9506 & 0.9111 \\
        \bottomrule
    \end{tabular}
\end{table}
In our experiments, we designed two variants: w/o Static, which removes the static analysis module, and w/o RAG, which eliminates the retrieval-augmented component. As shown in Table ~\ref{tab:performance_comparison}, compared to the complete LLM-BSCVM framework, the removal of the static analysis module led to a significant decline in F1-score, accuracy, and precision. This result indicates that static analysis effectively identifies common and easily detectable vulnerability patterns, such as reentrancy attacks and integer overflows, which can often be recognized through simple pattern matching. Consequently, the removal of this module resulted in degraded model performance.
\begin{table}[htbp]
    \centering
    \caption{Results of The Ablation Experiments on LLM-BSCVM}
    \label{tab:performance_comparison}
    \begin{tabular}{lcccc}
        \toprule
        Approach & F1 & Recall & Precision & Accuracy \\
        \midrule
        W/o Static & 0.8848 & 0.8497 & 0.9228 & 0.8858 \\
        W/o RAG & 0.8440 & 0.7541 & 0.9483 & 0.8561 \\
        LLM-BSCVM(W) & 0.9104 & 0.8743 & 0.9506 & 0.9111 \\
        \bottomrule
    \end{tabular}
\end{table}
Similarly, the w/o RAG variant also exhibited a performance decline. The RAG module dynamically retrieves knowledge from a vulnerability database, supplementing the model's knowledge gaps in vulnerability detection. This component is particularly beneficial when encountering previously unseen vulnerability types, as it allows the model to access the latest relevant data in real time, thereby enhancing both accuracy and robustness.

\begin{figure}
  \centering
  \includegraphics[width=1\linewidth]{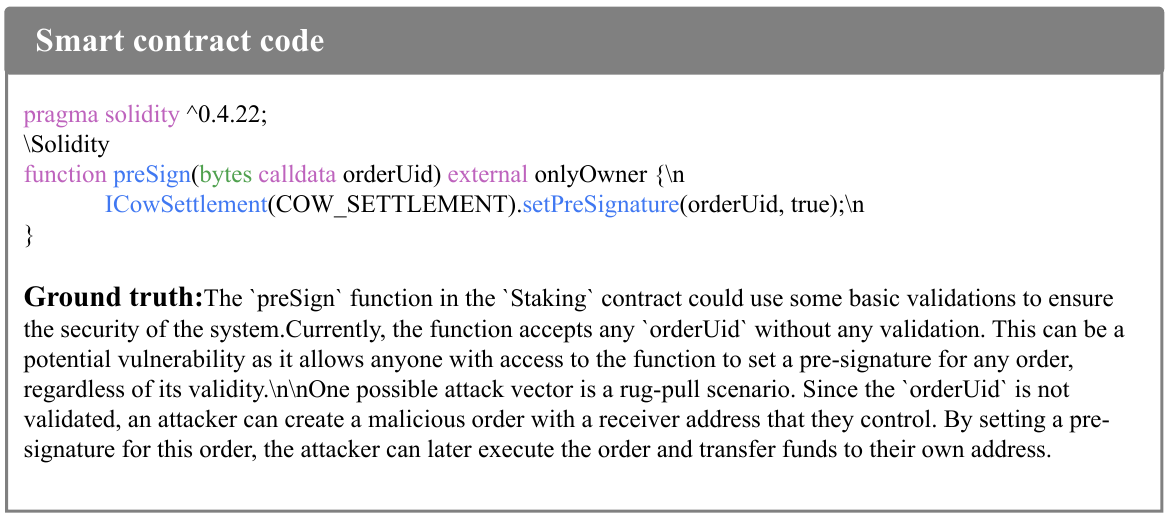}
  \caption{Example Contract: preSign Contract.}
  \label{fig:5}
\end{figure}

\indent\textbf{Other Tasks. }Figure \ref{fig:5} illustrates the preSign contract used to evaluate the effectiveness of LLM-BSCVM in four additional tasks: repair suggestion generation, risk level assessment, and vulnerability repair.

\begin{figure}
  \centering
  \includegraphics[width=1\linewidth]{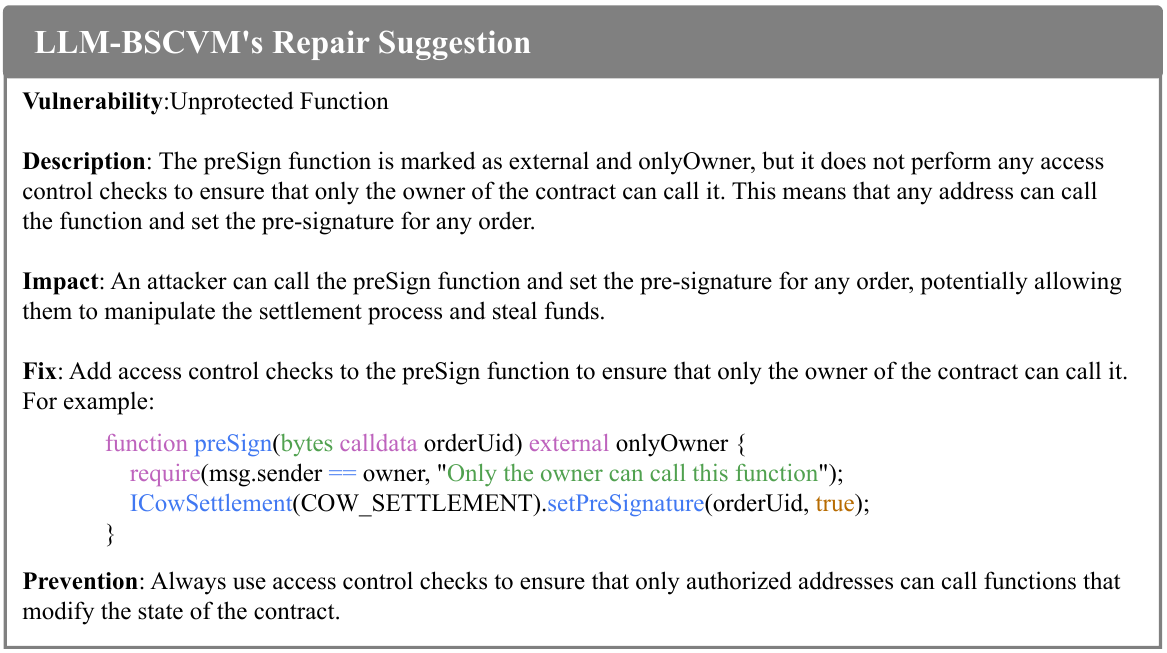}
  \caption{repair suggestion Results of LLM-BSCVM.}
  \label{fig:6}
\end{figure}

\begin{figure}
  \centering
  \includegraphics[width=1\linewidth]{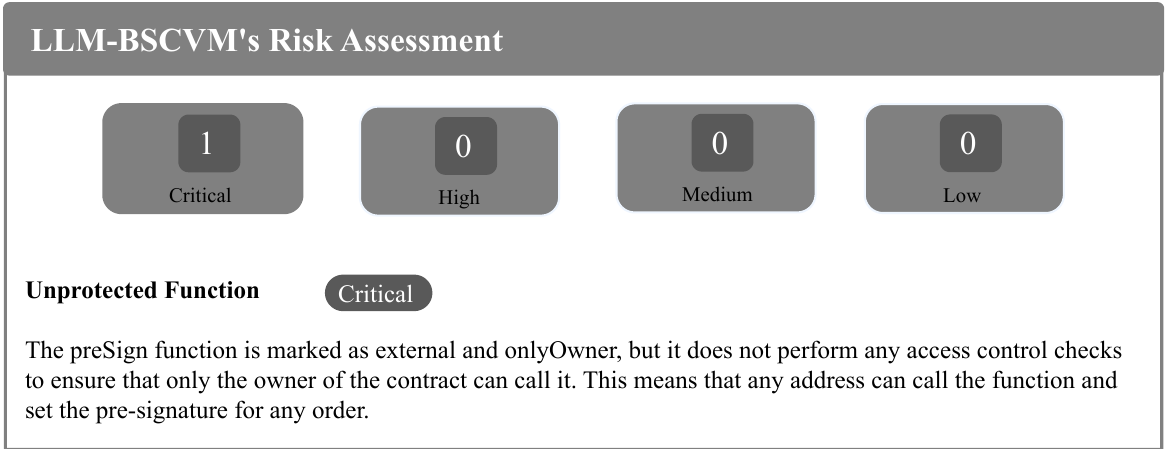}
  \caption{Risk Assessment Results of LLM-BSCVM.}
  \label{fig:7}
\end{figure}

\begin{figure}
  \centering
  \includegraphics[width=1\linewidth]{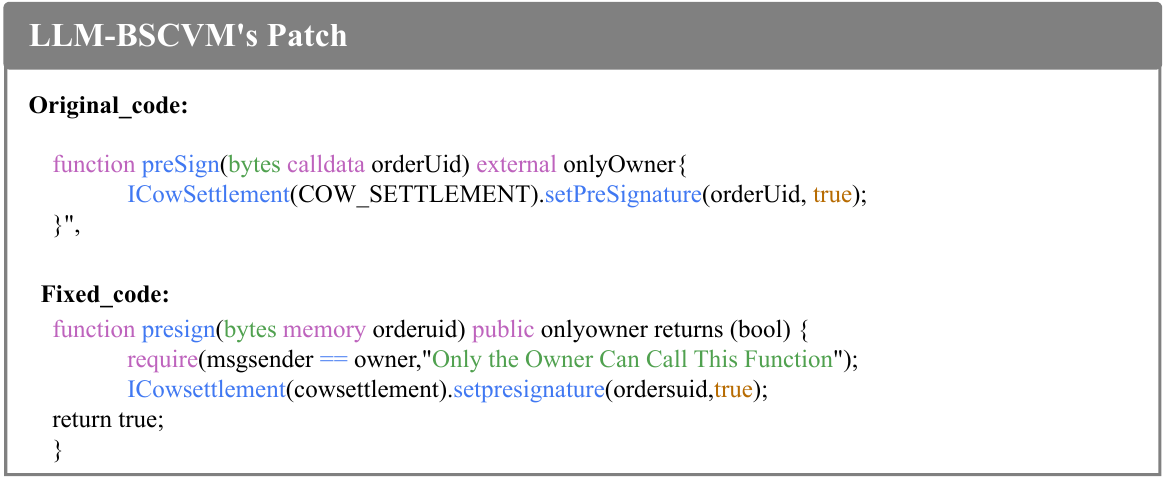}
  \caption{Vulnerability repair Results of LLM-BSCVM.}
  \label{fig:8}
\end{figure}

Figure \ref{fig:6} illustrates the repair suggestions generated by LLM-BSCVM, covering five aspects: vulnerability name, cause analysis, potential impact assessment, specific repair steps, and preventive measures recommendations. These suggestions provide a detailed description of the vulnerabilities in the preSign contract and their corresponding repair solutions. Compared to the ground truth repair suggestions, the suggestions generated by LLM-BSCVM exhibit high semantic consistency, both pointing out that an attacker could exploit the preSign function to set pre-signatures for any order, thereby leading to the risk of asset theft. Although the phrasing differs, LLM-BSCVM offers a comprehensive analysis of the vulnerability’s impact and presents the same essential repair solution.\\
\hspace*{\parindent}Figure \ref{fig:7} presents the risk level assessment results for the preSign contract. For the identified vulnerability, “Unprotected Function, " LLM-BSCVM correctly assigns a “Critical" risk level and provides a statistical distribution of different risk levels.\\
\hspace*{\parindent}Figure \ref{fig:8} displays the results after the vulnerability was repaired. The repaired contract introduces access control checks in the preSign function, ensuring that only the contract owner can invoke this function, along with the necessary validation logic, effectively eliminating the vulnerability.

\indent\textbf{Vulnerability repair. }
To evaluate the effectiveness of LLM-BSCVM in vulnerability repair, we utilized another independent large model to validate the repaired smart contracts. The results show that approximately 21\% of the contracts successfully passed the validation, indicating that the model is capable of vulnerability repair and can effectively reduce the impact of certain high-risk vulnerabilities. For the contracts that did not pass the validation, further analysis revealed that, during the repair process, some contracts might have adjusted their code structure or security mechanisms, leading to the detection of potential risks under certain evaluation criteria. Additionally, discrepancies in evaluation standards can lead to different validation results. Therefore, further optimization of the repair strategy is needed to improve the success rate of repairs.\\
\indent\textbf{Report Generation. }Although the automatically generated reports have not yet fully reached the reliability level of human expert reports, their advantages are significant. They not only provide detailed repair suggestions but also include the repaired contract code as a reference. Furthermore, compared to traditional manual audit reports, the generation time is significantly shortened, greatly improving efficiency.
\section{Conclusion}
The widespread application of smart contracts in the Web 3.0 ecosystem is accompanied by significant security challenges, where vulnerabilities can lead to substantial economic losses and systemic risks. To address this, this paper proposes LLM-BSCVM, the first end-to-end vulnerability management framework for smart contracts, designed to provide comprehensive functions for vulnerability detection, root cause analysis, repair recommendations, risk assessment, and audit reporting. The core innovation of LLM-BSCVM lies in its “Decompose-Retrieve-Generate" three-stage approach, which includes: (1) Task Decomposition, based on the concept of multi-agent collaboration, breaking down the vulnerability management process to facilitate progressive reasoning in vulnerability detection, repair suggestions, and risk assessment; (2) Knowledge Retrieval, integrating the vulnerability knowledge base with external data sources in real-time to enhance contextual understanding; (3) Result Generation, where agents combine retrieved relevant knowledge to generate explainable vulnerability analysis, repair plans, and final security audit reports. Experimental evaluation shows that LLM-BSCVM achieves a vulnerability detection accuracy and F1 score of 91\% on benchmark datasets, while the false positive rate decreases from the state-of-the-art (SOTA) 7.2\% to 5.1\%, enhancing the reliability and feasibility of vulnerability repair while maintaining high detection performance. Future work will focus on (1) integrating symbolic execution and formal verification to improve detection accuracy, and (2) optimizing LLM reasoning transparency through human-computer interaction. LLM-BSCVM is expected to advance the application of AI in smart contract security and provide more intelligent, automated security guarantees for the Web 3.0 ecosystem.


\end{document}